%Paper: gr-qc/9211001
%From: NARDELLI%ITNCISCA.bitnet@ICINECA.CINECA.IT
%Date: 03 Nov 1992 10:20:22 +0000
%Date (revised): 09 Nov 1992 09:06:18 +0000

\def\square{\kern1pt\vbox{\hrule height 1.2pt\hbox{\vrule width 1.2pt\hskip 3pt
   \vbox{\vskip 6pt}\hskip 3pt\vrule width 0.6pt}\hrule height 0.6pt}\kern1pt}
\magnification 1200

\hoffset=-.1in
\voffset=-.2in
\vsize=7.5in
\hsize=5.6in
\tolerance 10000

\baselineskip 12pt plus 1pt minus 1pt
\pageno=0
\centerline{\bf CHERN-SIMONS GRAVITY FROM 3+1-DIMENSIONAL GRAVITY}
\smallskip
%\centerline{\bf Titolo}
\smallskip
\vskip 24pt
\centerline{ G. Grignani}
\vskip 8pt
\centerline{\it Dipartimento di Fisica}
\centerline{\it Universit\'a degli Studi di Perugia}
\centerline{\it I-06100  Perugia -- ITALY}
\vskip 12pt
\centerline{and}
\vskip 12pt
\centerline{ G. Nardelli}
\vskip 8pt
\centerline{\it Dipartimento di Fisica}
\centerline{\it Universit\'a degli Studi di Trento}
\centerline{\it I-38050  Povo (TN) -- ITALY}

\vskip 1.5in
\centerline{Submitted to: {\it Physics Letters B}}
\vfill
\vskip -12pt
\noindent  DFUPG-59-1992

\noindent  UTF-272-1992
\hfill September 1992
\eject
\baselineskip 12pt plus 2pt minus 2pt

\centerline{\bf ABSTRACT}
\medskip
In the context of a Poincar\'e gauge theoretical formulation, pure gravity
in $3+1$-dimensions is dimensionally reduced to gravity in $2+1$-dimensions
with or without cosmological constant $\Lambda$. The dimensional
reductions are consistent with the gauge symmetries, mapping $ISO(3, 1)$
gauge transformations into $ISO(2,1)$ ones. One of the reductions leads to
Chern-Simons-Witten gravity.

The solutions of $2+1$-gravity with $\Lambda\le 0$
(in particular the black-hole solution recently found by Banados,
Teitelboim and Zanelli) and
 those of   $1+1$-dimensional Liouville gravity, are thus
mapped into $3+1$-dimensional vacuum solutions.

\vfill
\eject

\medskip
\nobreak
$2+1$-dimensional gravity$^1$ has attracted a growing attention in the
last decade, both as a good theoretical laboratory for the construction of
a quantum theory of gravity$^{2,3}$ and as a simplified model to study
``physical'' gravitational systems$^4$. Such roles played by
$2+1$-dimensional gravity are well exemplified by
the black-hole solution of Einstein equations with negative
cosmological constant  recently found by  Banados et al.$^{5,6}$.
The key
properties of this solution are similar to
those of its $3+1$-dimensional counterpart (apart from not being
asymptotically flat), so that one can study black-hole physics in a much
simpler setting.

In this context it is then interesting to establish a clear link
between $D=4$ and $D=3$ gravities by a dimensional reduction.
 This is the aim of the present letter and it is
achieved in the framework of a gauge theoretical formulation of both
theories. In fact as we showed in Ref.[7] (see also [8]) gravity in three
and four dimensions with all possible couplings to matter fields and
particles can be formulated as a gauge theory of the Poincar\'e group. In
three  dimensions this formulation is especially attractive as the
Einstein-Hilbert action becomes the Chern-Simons term of the $ISO(2,1)$
gauge group$^{2,9}$.
Such a Chern-Simons action with the correct $ISO(2,1)$ gauge
transformations can then be derived by dimensionally reducing the
$3+1$-dimensional Einstein-Hilbert action in its gauge theoretical
formulation.

The same dimensional reduction process was used in Refs.[10] (see also
[11]) to obtain, starting from an $ISO(2,1)$ gauge invariant theory in
$2+1$-dimensions, the gauge theoretical formulations$^{12}$ of the
$1+1$-dimensional Liouville$^{13}$ and black-hole gravity$^{14}$.

As we provide the link between 4 and 3 dimensions, $1+1$-dimensional
solutions of Liouville and black-hole gravity can then be directly
connected to four dimensional solutions of Einstein's equations.

We shall perform two different dimensional reductions corresponding to two
different identifications of the $ISO(3,1)$ generators of the four
dimensional theory, to the $ISO(2,1)$ generators of the $2+1$-dimensional
one. Whereas the first reduction  leads to a theory with vanishing
cosmological constant, the second will produce a theory
in which a (negative) $3D$-cosmological constant is related to the
$4D$-Newton constant. Therefore the $3D$-black-hole solution of Ref.[5] will
be easily read as the cross section of a $4D$ ``black-string'' solution of
the Einstein equations in vacuo.

We shall not report here on the dimensional reduction from $4D$ to $3D$
gravity in the presence of matter fields and point-particles, because such
a reduction can be easily achieved following the lines of Ref.[10] for the
$3D\to 2D$ reduction and starting from the matter field actions provided in
Ref.[7].

We shall begin by briefly reviewing our approach to gravity as a gauge
theory of the Poincar\'e group and then we shall present the dimensional
reductions.

Key ingredients of the formulation of gravity as a Poincar\'e
gauge theory are the so
called Poincar\'e coordinates $q^a(x)$, Higgs type fields that behave as
vectors under Poincar\'e gauge transformations,$^{7,8}$ and that are
involved in the definition of the vielbein $V^a{}_\mu$. The $q^a$ can be
interpreted as coordinates of an internal Minkowskian space ${\cal M}_q$ that
ca
   n
be locally made to coincide with the tangent space.
Any choice of Poincar\'e coordinates is equivalent to a gauge choice
leaving the theory invariant under residual (local) Lorentz
transformations. In our formalism the vielbein $V^a{}_\mu$ is not
identified with the component $e^a{}_\mu$ of the gauge potential $A_\mu$
along the translation generators $P^a$,
 but it will be given by the Poincar\'e
covariant derivative of the coordinates $q^a$, namely $V^a{}_\mu={\cal D}_\mu
q^a = \partial_\mu q^a + \omega^a{}_{b{}\mu} q^b + e ^a{}_\mu$, where
$\omega^{ab}{}_\mu$ is the spin connection.
Only in the so called ``physical'' gauge, where $q^a = 0$, $V^a{}_\mu =
e^a{}_\mu$. But this interpretation for $e^a{}_\mu$
only holds in a particular gauge choice of the translations and, consequently,
in the  framework of a {\it Lorentz} gauge theory.

 For
a more detailed discussion on the Poincar\'e coordinates and on the
necessity of their introduction in the context of a Poincar\'e gauge theory
of gravity see Ref.[7,8]. It is important to know, however, that this
approach to gravity as a Poincar\'e gauge theory
allows to couple in a gauge invariant
fashion particle and matter fields to gravity in any dimensions,
obtaining equations that are equivalent to Einstein's
equation with  matter sources,  to field theory in curved space or to the
geodesic equations in the case of particles. Gravity becomes in this way as
close as possible to any ordinary non-Abelian gauge theory and, even if in
$4D$ the Einstein-Hilbert action does not assume the
Yang-Mills form, the gauge fields transform in the usual way.

The Lorentz and momentum generators\footnote{$^\dagger$}{In our
notations, latin indices $a, b, c, ...=
0,1,2$ and capital latin indices $A, B, C, ... = 0, 1, 2, 3$ denote
$ISO(2,1)$ and $ISO(3,1)$ internal (gauge) indices, respectively. They
are raised and lowered by the Minkowski metrics $\eta_{ab} = (1, -1, -1)$
and $\eta_{AB} = (1, -1, -1, -1)$. In the dimensional reduction the
first 3 values of $A, B, C, ...$ will denote the corresponding $ISO(2,1)$
internal indices $a, b, c, ... ,  i.e.  A =(a,3), B=(b,3), C=(c,3), ...$. The
$3+1$ dimensional space-time indices are denoted by the first greek letters
$\alpha, \beta, \gamma, ... = 0, 1, 2, 3$ whose first three components
denote the corresponding $2+1$-dimensional space-time indices $\mu, \nu,
\rho, ... =0, 1, 2$. We shall use the antisymmetric symbol
$\varepsilon^{ABCD}$ with $\varepsilon^{0123} = 1$ and in $2+1$-dimensions
$\varepsilon^{abc} = \varepsilon^{abc3}$, so that $\varepsilon^{012}=1$.}
$J_{AB}$ and $P_A$ satisfy the Poincar\'e algebra
$$\eqalign{\left[ P_A, P_B\right] &= 0\ \ ,\cr
\left[ P_A, J_{BC}\right] &= \eta_{AC} P_B - \eta_{AB}P_C\ \ ,\cr
\left[ J_{AB}, J_{CD}\right] &= \eta_{AC} J_{BD} - \eta_{BC} J_{AD} +
\eta_{BD}J_{AC} - \eta_{AD} J_{BC}\ \ .\cr}\eqno(1)$$
As we mentioned, in order to have gauge fields transforming in the usual way
under non-Abelian gauge transformations, the {\it vierbein} is defined through
the covariant derivative of a Poincar\'e vector $q^A$, namely of a quantity
that
under gauge transformations behaves as $$\delta q^A(x) = \kappa^A{}_B(x) q^B(x)
+ \rho^A(x)\ \ \ ,\eqno(2)$$ where $\kappa^{AB}=-\kappa^{BA}$ and $\rho^A$ are
the infinitesimal parameters corresponding to Lorentz transformations and
translations, respectively. The $ISO(3,1)$ covariant derivative of the
Poincar\'e coordinates  $q^A$ will contain an homogeneous part, with $SO(3,1)$
gauge potentials $\omega^{AB}{}_\alpha$, and an inhomogeneous part
with gauge potentials $e^A{}_\alpha$ associated to the
translation generators $P_A$, namely
$${\cal D}_\alpha q^A =
\partial_\alpha q^A +\omega^{AB}{}_\alpha q_B + e^A{}_\alpha\ \ .\eqno(3)$$

We need to construct, in terms of ${\cal D}_\alpha q^A= V^A{}_\alpha$, an
$ISO(3,1)$ gauge scalar that can serve as a metric on the space-time, namely a
quantity of the type $g_{\alpha\beta} =
\eta_{AB} V^A{}_\alpha  V^B{}_\beta =
\eta_{AB} {\cal D}_\alpha q^A {\cal D}_\beta q^B$ . For this purpose
${\cal D}_\alpha q^A$ has to transform as a Lorentz vector under the gauge
transformations, {\it i.e.} as $\delta {\cal D}_\alpha q^A = \kappa^A{}_B
{\cal D}_\alpha q^B$. As a consequence one has to impose that $e^A{}_\alpha$
and $\omega^{AB}{}_\alpha$ under gauge transformations change as
$$\eqalign{ \delta \omega^{AB}{}_\alpha &= -\partial_\alpha \kappa^{AB}
-\omega^{AC}{}_\alpha\kappa_C{}^B + \omega^{BC}{}_\alpha\kappa_C{}^A
\ \ ,\cr \delta e^{A}{}_\alpha & = -\partial_\alpha \rho^{A}
-\kappa^{A}{}_{B}e^B{}_\alpha - \omega^{AB}{}_\alpha\rho_B \ \ .\cr}
\eqno(4)$$
The covariant derivative (3) (as we showed in Ref.
[7] by gauging the action of a free relativistic particle) can then be
interpreted as the space-time {\it vierbein} $V^A_\alpha$.

The transformation laws (4), are those introduced by Witten$^2$ for the
$ISO(2,1)$ Chern-Simons formulation of gravity in $3D$. Consequently, in our
framework, the dimensional reductions that will be consistent with the
$ISO(2,1)$ gauge invariance in  $3D$, will naturally arise.

Introducing the Lie algebra
valued gauge potential $A_\alpha = e^A{}_\alpha P_A - (1/ 2)
 \omega^{AB}{}_\alpha J_{AB}$ and gauge parameter
$u = \rho^A P_A -(1/ 2)\kappa^{AB} J_{AB}$, the
transformation laws (4) become those of
 any ordinary non Abelian gauge theory, i.e.
$\delta A_\alpha = -\partial_\alpha u - [A_\alpha  , u] \equiv -
\Delta_\alpha u$ .

The Lie-algebra valued field strength is
$$ F_{\alpha \beta} = [\Delta_\alpha , \Delta_\beta] = P_A
T^A{}_{\alpha  \beta} - {1\over 2} J_{AB} R^{AB}_{\alpha \beta} \ \ \
,\eqno(5)$$
where
$$ \eqalign{ T^A_{\alpha\beta}& = \partial_\alpha e^A{}_\beta - \partial_\beta
e^A{}_\alpha +  \omega^{AB}{}_\alpha e_{B\beta} - \omega^{AB}{}_\beta
e_{B\alpha}\ \ , \cr
R^{AB}_{\alpha\beta} &= \partial_\alpha  \omega^{AB}{}_\beta - \partial_\beta
\omega^{AB}{}_\alpha + \omega^{AC}{}_\alpha \omega_{C\ \beta}^{\ B}
-\omega^{AC}{}_\beta\omega_{C\ \alpha}^{\ B} \  . \cr}\eqno(6)$$

$F_{\alpha\beta}$ transforms covariantly under gauge transformations and
whereas $R^{AB}_{\alpha\beta}$ can be interpreted as the Riemann curvature
tensor, $T^A_{\alpha\beta}$ does not correspond to the space-time
torsion, as the {\it vierbein} is not given by $e^A{}_\alpha$.
The space-time torsion ${\cal T}^A{}_{\alpha \beta}$ can be easily
evaluated in term of the $ISO(3,1)$ field strength and Poincar\'e
coordinates as
$$ \eqalign{{\cal T}^A_{\alpha\beta} &= \partial_\alpha V^A{}_\beta -
\partial_\beta V^A{}_\alpha +  \omega^{AB}{}_\alpha V_{B\beta} -
\omega^{AB}{}_\beta V_{B\alpha}\cr
&= T^A_{\alpha\beta} + R^{AB}_{\alpha\beta} q_B\ \ ,\cr} \eqno(7)$$
so that only in the physical gauge ${\cal T}^A_{\alpha\beta}=
T^A_{\alpha\beta}$.
 Within this formalism the Einstein-Hilbert action
$S^{4D}_{\rm EH} = ( 4\pi G)^{-1}\int d^4x\sqrt{-g} R$, where $R$ is the scalar
curvature and $G$ the Newton constant, can be rewritten in the form of a
$ISO(3,1)$ gauge invariant action according to
$$S^{4D}_{\rm EH} = -{1\over 16\pi G} \int d^4 x \, \varepsilon^{\alpha \beta
\gamma\delta} \varepsilon_{ABCD} {\cal D}{}_\alpha q^A
{\cal D}{}_\beta q^B R^{CD}_{\gamma\delta}\ \ . \eqno(8)$$
Had we used $e^A{}_\alpha$ instead of ${\cal D}{}_\alpha q^A $ the action
would not be Poincar\'e gauge invariant.

The equations of motion obtained
by varying (8) with respect to $e^A{}_\alpha$, $\omega^{AB}{}_\beta$ and
$q^A$ give, respectively, (provided that the {\it vierbein} is invertible)
the vanishing of the space-time torsion ${\cal T}^A_{\alpha\beta}$, the
Einstein equations in vacuo, and an equation that is automatically
satisfied if the other two are used.
In $2+1$ dimensions something peculiar happens.
Here in fact the Einstein-Hilbert action  $S^{3D}_{\rm EH} = (
4\pi G_{3D})^{-1}\int d^3 x\sqrt{g} R$, in its $ISO(2,1)$ gauge invariant form,
reads
$$S^{3D}_{\rm EH} = {1\over 8\pi G_{3D}} \int d^3 x \,
\varepsilon^{\mu\nu \rho} \varepsilon_{abc} {\cal D}{}_\mu q^a
R^{bc}_{\nu\rho}\ \ . \eqno(9)$$
and by means of the Bianchi identity
$\varepsilon^{\mu\nu\rho}D_\mu R^{bc}_{\nu\rho} = 0 $,
$S^{3D}_{\rm EH}$ becomes
$$S^{3D}_{\rm EH} = {1\over 8\pi G_{3D}} \int d^3 x \, \varepsilon^{\mu\nu
\rho} \varepsilon_{abc} e^a{}_\mu
R^{cd}_{\nu\rho}\ \ , \eqno(10)$$
up to a surface term that can always be chosen to vanish.
($G_{3D}$ in three dimensions has the mass dimension $[M]^{-1}$).
Therefore all the terms containing $q^a$ disappear from the action
$S^{3D}_{\rm EH}$ and as long as pure gravity is concerned,
 $e^a{}_\mu$ can indeed be interpreted as the space-time
{\it dreibein} and yet the theory is Poincar\'e gauge invariant, contrary
to what happens in $3+1$-dimensions.
The absence of the $q$ variables in (10) and the interpretation of $e$ and
$\omega$ as gauge fields makes of (10) an action of the form $AdA + A^3$
that can be conceived as a Chern-Simons three form. Introducing the dual of
the connection $\omega$ and of the Lorentz generators,
with a suitable non-degenerate and invariant inner product among the $ISO(2,1)$
generators, $S^{3D}_{\rm EH}$ actually becomes the $ISO(2,1)$
Chern-Simons three-form$^2$.

If matter or a gravitational constant is included and one still looks for a
Poincar\'e gauge theory, the $q^a$ variables have to be reintroduced. So
that for example in the presence of a cosmological constant $\Lambda_{3D}$,
the $ISO(2,1)$ gauge invariant action reads
$$S^{3D}_{\Lambda} =  \int d^3 x \, \varepsilon^{\mu\nu
\rho} \varepsilon_{abc}\left({1\over 8\pi G_{3D}} e^a{}_\mu
R^{bc}_{\nu\rho} - {\Lambda_{3D}\over 3!}{\cal D}{}_\mu q^a
{\cal D}{}_\nu q^b{\cal D}{}_\rho q^c\right)\ \ . \eqno(11)$$

We want to show that there are two suitable dimensional reductions of the
Poincar\'e generators of the $ISO(3,1)$ theory and correspondingly of the
space-time dimensions that from the action (8) and the algebra (1) lead to
the Poincar\'e gauge theories (10) and (11).
With such reductions from the $ISO(3,1)$ gauge transformations (2) and (4), we
shall obtain the corresponding gauge transformations in $2+1$-dimensions.

We shall eliminate in both cases the third spatial dimension (of coordinate
$x^3$) that will be compactified to a unit length. Furthermore, we shall set
$\partial_3 (anything) =0 $ so that the  integral in $x^3$ in (8) will
only give an overall unit factor.
The dimensional reduction leading from the $4$-dimensional  $ISO(3,1)$
theory of Eq.(8) to $ISO(2,1)$ Chern-Simons gravity is given in Table A.

\bigskip
\bigskip
\centerline{\it TABLE A}
\bigskip

\centerline{
\vbox{ \offinterlineskip \hrule
\def\tablerule{\noalign{\hrule}}
\halign{\vrule#&\strut\quad#\hfil\quad&
\vrule#&\quad\strut\hfil#\quad&\vrule#\cr
height4pt&\omit&&\omit&\cr
&\multispan3\hfil {\bf Dimensional Reduction A}\hfil &\cr
height4pt&\omit&&\omit&\cr
height4pt&\omit&&\omit&\cr \tablerule
height4pt&\omit&&\omit&\cr
&\hfil 3+1 Dimensions \hfil&& \hfil 2+1 Dimensions \hfil&\cr
height4pt&\omit&&\omit&\cr\tablerule
height4pt&\omit&&\omit&\cr
&\hfil $e^3{}_3$ \hfil&& \hfil $1$ \hfil&\cr
height4pt&\omit&&\omit&\cr\tablerule
height4pt&\omit&&\omit&\cr
&\hfil $e^a{}_\mu$\hfil &&\hfil $e^a{}_\mu$\hfil&\cr
height4pt&\omit&&\omit&\cr\tablerule
height4pt&\omit&&\omit&\cr
&\hfil $e^a{}_3$\hfil &&\hfil $0$\hfil&\cr
height4pt&\omit&&\omit&\cr\tablerule
height4pt&\omit&&\omit&\cr
&\hfil $e^3{}_\mu$\hfil &&\hfil $0$\hfil&\cr
height4pt&\omit&&\omit&\cr\tablerule
height4pt&\omit&&\omit&\cr
&\hfil$\omega^{ab}{}_\mu$ \hfil&& \hfil$\omega^{ab}{}_\mu$\hfil&\cr
height4pt&\omit&&\omit&\cr\tablerule
height4pt&\omit&&\omit&\cr
&\hfil$\omega^{a3}{}_\mu$ \hfil&& \hfil$0$\hfil&\cr
height4pt&\omit&&\omit&\cr\tablerule
height4pt&\omit&&\omit&\cr
&\hfil$\omega^{AB}{}_3$\hfil && \hfil$0$\hfil&\cr
height4pt&\omit&&\omit&\cr\tablerule
height4pt&\omit&&\omit&\cr
&\hfil$q^{a}$\hfil && \hfil$q^a$\hfil&\cr
height4pt&\omit&&\omit&\cr\tablerule
height4pt&\omit&&\omit&\cr
&\hfil$q^{3}$\hfil && \hfil$0$\hfil&\cr
height4pt&\omit&&\omit&\cr\tablerule
height4pt&\omit&&\omit&\cr
&\hfil$\rho^{a}$\hfil && \hfil$\rho^a$\hfil&\cr
height4pt&\omit&&\omit&\cr\tablerule
height4pt&\omit&&\omit&\cr
&\hfil$\rho^{3}$\hfil && \hfil$0$\hfil&\cr
height4pt&\omit&&\omit&\cr\tablerule
height4pt&\omit&&\omit&\cr
&\hfil$\kappa^{ab}$\hfil && \hfil$\kappa^{ab}$\hfil&\cr
height4pt&\omit&&\omit&\cr\tablerule
height4pt&\omit&&\omit&\cr
&\hfil$\kappa^{a3}$\hfil && \hfil$0$\hfil&\cr
height4pt&\omit&&\omit&\cr\tablerule}}}
\bigskip
\bigskip

One can check that the $ISO(3,1)$ gauge transformations (2), (4), with
the identifications of Table A are mapped respectively into
$$\eqalign{\delta q^a &= \kappa^a{}_b q^b + \rho^a\ \ \cr
\delta e^a{}_\mu &= -\partial_\mu\rho^a - \kappa^a{}_b e^b{}_\mu
-\omega^{ab}{}_\mu\,\rho_b\ \ ,\cr
\delta\omega^{ab}{}_\mu &= -\partial_\mu\kappa^a
+\kappa^a{}_c\omega^{cb}{}_\mu - \kappa^b{}_c\omega^{ca}{}_\mu\ \
,\cr}\eqno(12)$$
{\it i.e.} into the correct $ISO(2,1)$
gauge transformations. In particular
the quantities that are set to a constant in Table A consistently have
vanishing gauge transformations. By substituting the content of Table A
into the action (8) one gets
$$S^{4D}_{\rm EH}\rightarrow S^{3D}_{\rm EH} = {1\over 8\pi G_{3D}} \int d^3
x \, \varepsilon^{\mu\nu\rho} \varepsilon_{abc} e^a{}_\mu \left(
\partial_\nu\omega^{bc}{}_\rho -\partial_\rho\omega^{bc}{}_\nu
+\omega^b{}_{d\nu}\omega^{dc}{}_\rho -
\omega^c{}_{d\nu}\omega^{db}{}_\rho\right)\ \
, \eqno(13)$$
where the constant $(G_{3D})^{-1}= ( G)^{-1}\int dx^3$ is
positive and has the correct mass dimensions $[M]^{-1}$.
The right hand side of (13) is precisely $S^{3D}_{EH}$ given in Eq. (10).

 Table A provides
the most natural reduction induced by the compactification of the third
spatial dimension. In fact in the internal space we retain only the
generators of the translations in time and in the direction 1 and 2
($\rho^a\ne 0$ and $\rho^3=0$) and the only possible Lorentz
transformations: boost and rotations on the plane 1,2 whose generators are
$J_{ab}$ ($\kappa^{ab}\ne 0$ and $\kappa^{a3} = 0$).

Denoting by $\hat P_A$ and $\hat J_{AB}$ the $ISO(3,1)$ generators (we
introduce the hat in order to avoid confusion with the corresponding
$ISO(2,1)$ generators), the $ISO(2,1)$ generators that we get from the
dimensional reduction in Table A are $ P_a=\hat P_a$ and $ J_{ab} = \hat
J_{ab}$.

Also $S^{3D}_\Lambda$ can be obtained with the dimensional reduction of
Table A from the $4$-dimensional $ISO(3,1)$ gauge invariant action with
non-vanishing cosmological constant
$$S^{4D}_{\Lambda} =  - \int d^4 x \, \varepsilon^{\alpha\beta\gamma\delta}
\varepsilon_{ABCD}\left({1\over 16\pi G} {\cal D}_\alpha q^A
{\cal D}_\beta q^B
R^{CD}_{\gamma\delta} - {\Lambda\over 4!}{\cal D}_\alpha q^A {\cal
D}_\beta q^B{\cal D}_\gamma q^C{\cal D}_\delta q^D\right)\ \ . \eqno(14)$$

An analogous dimensional reduction has been used in Ref.[10] to obtain, from
$S^{3D}_\Lambda$,  the Poincar\'e gauge theoretical formulation of
Liouville gravity in $2$-dimensions. Thus the connection from the
$4$-dimensional Poincar\'e gauge theory (14) to the $2$-dimensional one is
established.

The second dimensional reduction we shall be concerned with, allows to
connect pure gravity in $4D$, Eq.(8), to $2+1$ gravity with a negative
cosmological constant, Eq.(11). The cosmological constant, will be
related to the $4D$ Newton constant through $(\Lambda_{3D}/ 3!) = -
( 4\pi G^2)^{-1}\int dx^3 = -(4 \pi G G_{3D})^{-1}$ . The dimensional
reduction is shown in Table B.

\bigskip
\bigskip
\centerline{\it TABLE B}
\bigskip
\centerline{
\vbox{ \offinterlineskip \hrule
\def\tablerule{\noalign{\hrule}}
\halign{\vrule#&\strut\quad#\hfil\quad&
\vrule#&\quad\strut\hfil#\quad&\vrule#\cr
height4pt&\omit&&\omit&\cr
&\multispan3\hfil {\bf Dimensional Reduction B}\hfil &\cr
height4pt&\omit&&\omit&\cr
height4pt&\omit&&\omit&\cr \tablerule
height4pt&\omit&&\omit&\cr
&\hfil 3+1 Dimensions \hfil&& \hfil 2+1 Dimensions \hfil&\cr
height4pt&\omit&&\omit&\cr\tablerule
height4pt&\omit&&\omit&\cr
&\hfil $e^3{}_3$ \hfil&& \hfil $1$ \hfil&\cr
height4pt&\omit&&\omit&\cr\tablerule
height4pt&\omit&&\omit&\cr
&\hfil $e^A{}_\mu$\hfil &&\hfil $0$\hfil&\cr
height4pt&\omit&&\omit&\cr\tablerule
height4pt&\omit&&\omit&\cr
&\hfil $e^a{}_3$\hfil &&\hfil $0$\hfil&\cr
height4pt&\omit&&\omit&\cr\tablerule
height4pt&\omit&&\omit&\cr
&\hfil$\omega^{ab}{}_\mu$ \hfil&& \hfil$\omega^{ab}{}_\mu$\hfil&\cr
height4pt&\omit&&\omit&\cr\tablerule
height4pt&\omit&&\omit&\cr
&\hfil$\omega^{a3}{}_\mu$ \hfil&& \hfil$ (G)^{-1/2} V^a{}_\mu
$\hfil&\cr height4pt&\omit&&\omit&\cr\tablerule
height4pt&\omit&&\omit&\cr
&\hfil$\omega^{AB}{}_3$\hfil && \hfil$0$\hfil&\cr
height4pt&\omit&&\omit&\cr\tablerule
height4pt&\omit&&\omit&\cr
&\hfil$q^{a}$\hfil && \hfil$0$\hfil&\cr
height4pt&\omit&&\omit&\cr\tablerule
height4pt&\omit&&\omit&\cr
&\hfil$q^{3}$\hfil && \hfil$\sqrt{G}$\hfil&\cr
height4pt&\omit&&\omit&\cr\tablerule
height4pt&\omit&&\omit&\cr
&\hfil$\rho^{A}$\hfil && \hfil$0$\hfil&\cr
height4pt&\omit&&\omit&\cr\tablerule
height4pt&\omit&&\omit&\cr
&\hfil$\kappa^{ab}$\hfil && \hfil$\kappa^{ab}$\hfil&\cr
height4pt&\omit&&\omit&\cr\tablerule
height4pt&\omit&&\omit&\cr
&\hfil$\kappa^{a3}$\hfil && \hfil$0$\hfil&\cr
height4pt&\omit&&\omit&\cr\tablerule}}}
\bigskip
\bigskip

The action $S^{4D}_{\rm EH}$ becomes
$$S^{4D}_{\rm EH}\rightarrow S^{3D}_{\Lambda} =  \int d^3 x \,
\varepsilon^{\mu\nu \rho} \varepsilon_{abc}\left({1\over 8\pi G_{3D}}
V^a{}_\mu R^{bc}_{\nu\rho} - {\Lambda_{3D}\over 3!} V^a{}_\mu
V^b{}_\nu V^c{}_\rho \right)\ \ , \eqno(15)$$
that is equivalent to Eq.(11) once the identification for the {\it
dreibein} $V^a{}_\mu = {\cal D}_\mu q^a$ and the Bianchi identity
 are taken into account.

The dimensional reduction B  leads from the $ISO(3,1)$ theory with generators
$(\hat P^A, \hat J^{AB})$ to a Lorentz, $SO(2,1)$ theory with generators
$J_{ab}$ in $2+1$ dimensions by setting to zero the generators $\hat P^A$ and
$\hat J^{a3}$ ($\rho^A= 0$ and $\kappa^{a3}=0$) and by identifying
$\hat J_{ab} = J_{ab}$ . The Lorentz theory is then transformed into a
Poincar\'e one by the definition of the dreibein $V^a{}_\mu$ as $V^a{}_\mu=
{\cal D}_\mu q^a$.

It is remarkable that only the theory with a negative cosmological constant
can be obtained , this is in fact the theory where the $2+1$-dimensional
black hole solution was found.
Such metric can be directly translated into a $3+1$-dimensional solution
with cylindrical symmetry of Einstein's equations in vacuo.

The 4-dimensional $ vierbein$ given by  $V^a{}_\mu=
{\cal D}_\mu q^a$ can be easily obtained from the $3$-dimensional one
using the identification of  Table B. Consequently the $4D$ vacuum solution
corresponding to the black-hole metric in $3D$ is given  by a line element
that is the same of the $3$-dimensional one with the addition of a $dz^2$
term with coefficient $g_{33}= -1$ and the cosmological constant
substituted by the Newton constant (with the suitable power to maintain mass
dimensions).
In a similar way , using the results of Ref.[10], one can connect the
Liouville gravity solutions in $2D$  to vacuum solutions in $4D$.

The dimensional reduction we have illustrated can be performed also in the
presence of matter fields and point-particles that can be coupled to
$ISO(3,1)$ gravity  in a gauge invariant fashion as in Ref.[7].
Such
descent does not entail any special difficulty and can be realized along
the line of Ref.[10]. In particular the solutions with fields and particles
presented in Ref.[10] for Liouville gravity can again be read as solutions
with fields and particles of gravity in $4D$ with $\Lambda=0$.

The Poincar\'e gauge theoretical formulation of gravitational theories is a
natural context to connect, preserving gauge invariance, such theories in
different dimensions. The dimensional descent provides a confirm that the
program of realizing gravity as a gauge theory of the Poincar\'e group
[7,15] can give interesting insights on the structure of the Einstein's
equations and on their solutions. At the same time it shows that what we
proposed in Ref.[7,10] is the natural generalization, to any
dimensions and with any matter couplings, of Witten's approach
to pure gravity in $2+1$-dimensions as a $ISO(2,1)$ gauge theory.

The solutions of $2+1$-dimensional gravity with a negative or vanishing
cosmological constant $\Lambda$ (in particular the $2+1$-dimensional
black-hole), and those of Liouville gravity, can be directly interpreted
 as vacuum  solutions of Einstein's
equation in $4D$. The study of their four dimensional properties and their
physical relevance is under investigation.

\vfill
\eject

\noindent{\bf REFERENCES}
\medskip
\nobreak
\bigskip

\item{[1]} S. Deser, R. Jackiw and G. t'Hooft, {\it Ann. Phys.} {\bf 152}
220 (1984); S. Giddings, J. Abbott and K. Kuchar, {\it Gen. Rel. Grav.}
{\bf 16} 751 (1984).
 \medskip
\item{[2]}  E. Witten, {\it Nucl. Phys.} {\bf B311}, 46 (1988).
\medskip
\item{[3]} A. Hosoya and K. Nakao, {\it Class. Quantum. Grav.} {\bf 7} 163
(1990); S. Carlip, {\it Phys. Rev.}, {\bf B42}, 2647 (1990).
\medskip
\item{[4]} See for example, S. Deser and R. Jackiw, {\it Time travel?},
BRX-TH 334, MIT-CTP\#2101 preprint, June (1992), and references therein.
\medskip
 \item{[5]} M. Banados, C. Teitelboim and J. Zanelli, IAS preprint,
IASSNS-HEP, 92/2.
 \medskip
\item{[6]} S. F. Ross and R. B. Maan, Waterloo preprint, WATPHYS-TH 92/07,
AUGUST 1992.
 \medskip
\item{[7]} G. Grignani and G. Nardelli, {\it Phys. Rev.} {\bf D45}, 2719
(1992).
\medskip
\item{[8]} K.S. Stelle and P.C. West, {\it Phys. Rev.} {\bf D21} 1466 (1980);
 W. Drechsler, {\it Ann. Inst. H.
Poincar\'e}, {\bf A XXXVII} 155 (1982);
T. Kawai, {\it Gen. Rel. Grav.}, {\bf 18}, 995 (1986) and in {\it The Sixth
Marcel Grossman Meeting on General Relativity}, H. Sato ed. (World
Scientific, Singapore, 1992).
\medskip
\item{[9]} A. Ach\'ucarro and P.Townsend, {\it Phys. Lett.} {\bf B 180},  85
(1986).
\medskip
\item{[10]} G. Grignani and G. Naredlli, Perugia-Trento preprint DFUPG
57/92, UTF-266-92, August 92.
 \medskip
\item{[11]}  A. Ach\'ucarro, Tufts University preprint, August (1992);
D. Cangemi, MIT Preprint CTP\# 2124, July 1992.
\medskip
\item{[12]}H. Verlinde, in {\it The Sixth Marcel Grossman Meeting on
General Relativity}, H. Sato, ed. (World Scientific, Singapore, 1992);
D. Cangemi and R. Jackiw, {\it
Phys. Rev. Lett.}, {\bf 69}, 233 (1992); R. Jackiw, in {\it
Recent Problems in  Mathematical Physics}, to be published in {\it Theor.
Math. Phys.}.
\medskip
\item{[13]} C. Teitelboim, {\it Phys. Lett.} {\bf B126}, 41 (1983), and
in {\it Quantum Theory of Gravity}, S. Christensen, ed. (Adam Higler,
Bristol, 1984);
R. Jackiw in {\it Quantum Theory of Gravity}, S. Christensen, ed. (Adam Higler,
Bristol, 1984) and {\it Nucl. Phys.} {\bf B252}, 343 (1985).
\medskip
\item{[14]} E. Witten, {\it Phys. Rev.} {\bf D44}, 314 (1991);
C. Callan, S. Giddings, A. Harvey and A. Strominger, {\it
Phys. Rev.} {\bf D45}, 1005 (1992); S. W. Hawking, {\it Phys. Rev. Lett.}
{\bf 69} 406 (1992).
 \medskip
\item{[15]}  G. Grignani and G. Nardelli, {\it Phys. Lett.} {\bf B264}, 45
(1991); G. Grignani and G. Nardelli, {\it Nucl. Phys.} {\bf B370}, 491
(1992).
\vfill
\eject

\end